\documentstyle[preprint,aps]{revtex}

\begin{document}

\title{Single j IBA}

\author{Shadow J.Q. Robinson$^{1}$ and Larry Zamick$^{2}$ }

\address{ 1) Department of Physics,\\
University of Southern Indiana, \\Evansville, Indiana  47712\\
2) Department of Physics and Astronomy,\\
Rutgers University, \\Piscataway, New Jersey  08854-8019}
\maketitle

\bigskip

In this report we examine two topics relating to previous
work. We feel that there are points to be made which we have
not made before. A common thread in the two problems is that they both
involve the isospin variable in an important way. 

In a publication by
Devi et. al.~\cite{devi1} we calculated the excitation energies of
T=T$_{min}$+1 states in odd A nuclei and of T=T$_{min}$+2 states of
even-even nuclei in the f-p shell where
T$_{min}$=$\frac{|N-Z|}{2}$. We performed a linear fit to these
excitation energies
\begin{eqnarray}
E(SA)=b(T+X) \nonumber \\
E(DA)=2b(T+X+\frac{1}{2}) \nonumber
\end{eqnarray}
For a simple interaction of the form a+b
t(1)$\cdot$t(2), the value of X is unity.

We point out that with a simple adjustment we can also convert this to
a formula for binding energies and hence obtain a term linear in
isospin, which was first found by Wigner.~\cite{wigner1,wigner2} We
simply assume that the binding energy is given by
\begin{equation}
E=\frac{b}{2}T(T+Y) \nonumber
\end{equation}
It is then easy to show that Y=2X-1. For the t(1)$\cdot$t(2)
interaction we have X=1, Y=1. For the Wigner SU(4) limit, we have
X=2.5, Y=4. It is worthwhile to note that in mean field theories we
cannot obtain a linear term in T, but in shell
model calculations it is impossible to not get such a term.

In~\cite{devi1} we performed a fit to the single j shell
calculation. We found that a good fit was obtained with b=2.32 and
X=1.3. This leads to a binding formula in the single j shell
\begin{equation}
E=\frac{b}{2}T(T+1.6) \nonumber
\end{equation}

In Talmi's book~\cite{talmi1}, expressions for the binding energy in
both the SU(4) limit and the seniority conserving limit are shown. In
the former case the binding energy goes as T(T+4) and in the latter as
T(T+1)~\cite{talmip}. It has been pointed out by McCullen
et. al.~\cite{mccullen1} that although seniority may be a pretty good
quantum number for a system of identical nucleons, e.g. the Calcium
isotopes, seniority is badly broken when we have both protons and
neutrons in open shells. This formula T(T+1.6) lies in between the two
extremes - one of seniority conservation for mixed protons and
neutrons and the other of the SU(4) limit in spin and isospin
variables.

The next problem we consider takes note of the fact that the ground
state wavefunctions of even-even Ti isotopes bear some resemblance to
IBA wavefunctions. There are various versions of the Interacting Boson
Approximation, IBA1~\cite{iba1}, IBA2~\cite{iba2}, and
IBA3~\cite{iba3}. The format of the Ti wavefunctions in
MBZ~\cite{mccullen1} most closely resembles that of IBA2.We here
define a model which we refer to as single j IBA.

In Table 1 we show the MBZ wavefunctions for the
J=0$^{+}$ T$_{min}$ ground states of $^{44,46,48}$Ti as well as the
unique (in the single j shell model)T$_{min}$+2 states. The
wavefunction is written as
\begin{equation}
\psi^{I}=\Sigma D^{I}(J_{p}J_{n})[(j^2)^{J_{p}}(j^r)^{J_n}]^{I} \nonumber
\end{equation}
where I is the total angular momentum and $D^{I}(J_{p}J_{n}$) is the
probability amplitude that the protons couple to J$_{p}$ and the
neutrons couple to J$_n$. For I=0 J$_{p}$= J$_{n}$=J.

In the single j model space, the states with the higher isospin
T$_{min}$+2 are not affected by any isospin conserving two nucleon
interaction. In fact for these states the coefficients
$D^{I}(J_{p}J_{n}$) are two particle coefficients of fractional
parentage. The reason for this is that these states in Ti are double
analogs of corresponding states in Ca, and for Ca we are dealing with
a system of identical particles i.e. only f$_{7/2}$ neutrons. A two
particle cfp will be an expansion in which neutrons are separated into
(n-2) and 2. We can then easily see the following for I=0
\begin{equation}
D^{I}(JJ)=(j^{n}J;j^2 J|\}j^{n+2}0) \nonumber
\end{equation}
And as the $D^{I}(J_pJ_n)$ satisfy the orthonormality conditions
\begin{equation}
\Sigma_{J_nJ_p}D^{I \alpha}(J_pJ_n)D^{I \alpha '}(J_pJ_n) =
\delta_{\alpha \alpha'} \nonumber
\end{equation}
So that in particular any T$_{min}$ state is orthogonal to a state
with T=T$_{min}$+2

We now define the single j IBA. Note that the ground state
wavefunction amplitudes for all three Ti isotopes. The largest
amplitudes have J$_p$=J$_n$=0 and J$_p$=J$_n$=2. The other amplitudes
ie (4,4), (6,6) etc are very small. This motivates us to consider a
simple model where the only non-vanishing D's are D(00) and D(22).
\begin{equation}
\psi \approx D(00)[(j^2)^0(j^n)^0]^0+D(22)[(j^2)^2(j^n)^2]^0 \nonumber
\end{equation}
With the conditions that 
\begin{equation}
D(00)^{2}+D(22)^{2}=1 \nonumber
\end{equation}
and that the orthonormality to the T$_{min}$+2 state is maintained
\begin{equation}
D(00)(j^n0;j^20|\}j^{n+2}0) + D(22)(j^n2;j^22|\}j^{n+2}0)=0 \nonumber
\end{equation}
But these two conditions mean that D(00) and D(22) are completely
determined - there is no freedom. We can show that the wavefunctions,
written as two component vectors for the various Ti isotopes are
\begin{eqnarray}
\psi_{^{44}Ti}=\frac{1}{\sqrt{14}}(\sqrt{5},3)=(0.5976,0.8018)
\nonumber \\
\psi_{^{46}Ti}=\frac{1}{\sqrt{8}}(\sqrt{5},\sqrt{3})=(0.7906,0.6124)
\nonumber \\
\psi_{^{48}Ti}=\frac{1}{\sqrt{6}}(\sqrt{5},1)=(0.9129,0.4082)
\nonumber
\end{eqnarray}

This comes from a more general expression~\cite{zml1,zml2} of Zamick,
Mekjian, and Lee.
\begin{eqnarray}
 \sqrt{\frac{(2j+1-n)}{(n+1)(2j+1)}} D(00)
            - M \sqrt{\frac{2n}{(n+1)(2j+1)(2j-1)}}
      &=& \left\{\matrix{0 & T = T_{min} \cr 1 & T = T_{min}+2} \right.
    \label{d00m}
\end{eqnarray}
where M=$\Sigma_{J \ge 2} D(JJ) \sqrt{(2J+1)}$.

Comparing with the results of Table 1 we see that for $^{44}$Ti there
is too much J=2 coupling - more than J=0. However the trend as one
goes through the Ti isotopes is quite reasonable and the wavefunctions
for $^{48}$Ti are remarkably similar. 

Toe coefficients D(22) play an important role in the calculations of
M1 transitions in the single j shell. The expression for
B(M1)$\uparrow$ from a J=$0^+$ to J=$1^+$ in units of $\mu_{N}^{2}$ is
given by~\cite{zamick1}
\begin{equation}
B(M1)=\frac{3}{4 \pi} (g_p - g_n)^{2} | \Sigma_{JV} D^{0}(J, J_{V})
D^{1}(J, J_V) \sqrt{J(J+1)}|^{2}
\end{equation}
Here g$_p$ and g$_n$ are the Schmidt values. If we sum over all
J=1$^{+}$ final states we obtain~\cite{zamick2} 
\begin{equation}
\Sigma_{\alpha} B(M1) =\frac{3}{4 \pi} (g_p - g_n)^{2} [ \Sigma_{J}
D^0 (J,J)^2  J(J+1)]
\end{equation}
If we only allow up to J=2 coupling then
\begin{equation}
B(M1)= \frac{18}{4 \pi} (g_p - g_n)^{2} |D^0(2,2) D^1(2,2)|^2
\end{equation}

%

Here we make a comparison of the MBZ and singlej IBA for the summed
strengths, using the effective value for $(g_p-g_n)=1.89$ as in
~\cite{zml2}.

\begin{tabular}{ccc}
$\Sigma$B(M1) & MBZ & singlej IBA\\
$^{44}$Ti     & 2.881 & 3.289 \\
$^{46}$Ti     & 1.919 & 1.9767\\
$^{48}$Ti     & 0.8528 & 0.8568\\
\end{tabular}

The values for $^{46}$Ti and $^{48}$Ti are remarkably similiar for MBZ
and singlej IBAeven though the values of D(22) are quite different. It
appears that the higher J contributions conspuire to make the summed
B(M1)'s for MBZ about the same as for singlej IBA.

The important point we wish to make here is that in the single j shell
model, once we make the assumption that the T=T$_{min}$ state consists
of only J$_{p}$=J$_{n}$=0 and J$_{p}$=J$_{n}$=2 couplings, the
relative amounts of the couplings is fixed. There is no freedom. The
reason for this is that the states with T$_{min}$ must be orthogonal
to the states with T$_{min}$+2. A small amount of the higher J
couplings restores the freedom to adjust the relative amounts of J=0
and J=2.

This work was supported by the U.S. Dept. of Energy under Grant
No. DOE-FG01-04ER04-02.

\begin{table}
\caption{Wave functions of $I=0_1$, $T_{min}$ and $I=0$, $T_{min} +2$
states of $^{44}$Ti, $^{46}$Ti and $^{48}$Ti. The symbol $*$ means $v
= 4$.$^{a}$}
\begin{tabular}{rcccc}
 $^{44}$Ti & $J_P$  &  $J_N$  &  $I=0$ $T=0$   &   $I=0$ $T=2$  \\
           &   0    &    0    &    0.7608      &   0.5000       \\
           &   2    &    2    &    0.6090      &  --0.3727  \    \\
           &   4    &    4    &    0.2093      &  --0.5000  \    \\
           &   6    &    6    &    0.0812      &  --0.6009  \    \\
  \hline
 $^{46}$Ti & $J_P$  &  $J_N$  &  $I=0$ $T=1$   &   $I=0$ $T=3$  \\
           &   0    &    0    &    0.8224      &   0.3162      \\
           &   2    &    2    &    0.5420      &  --0.4082 \     \\
           &   2    &  2$^*$  &    0.0563      &   0.0          \\
           &   4    &    4    &    0.0861      &  --0.5477 \     \\
           &   4    &  4$^*$  &  --0.1383 \    &   0.0          \\
           &   6    &    6    &  --0.0127 \    &  --0.6583 \     \\
  \hline
 $^{48}$Ti & $J_P$  &  $J_N$  &  $I=0$ $T=2$   &   $I=0$ $T=4$  \\
           &   0    &    0    &    0.9136      &   0.1890       \\
           &   2    &    2    &    0.4058      &  --0.4226 \     \\
           &   4    &    4    &    0.0196      &  --0.5669 \     \\
           &   6    &    6    &  --0.0146 \    &  --0.6814 \      
\end{tabular}
a)The phases have been adjusted to fit with the cfp conventions
of~\cite{talmi1} and differ in some way with those
in~\cite{mccullen1}.
\end{table}


\begin{thebibliography}{99}

\bibitem{devi1} Y.Durga Devi,
S.J.Q. Robinson and L. Zamick, Phys. Rev. C {\bf 61}, 037305 (2000).

\bibitem{wigner1} E. Wigner, Phys. Rev. \textbf{51}, 106 (1937).
 
\bibitem{wigner2} E. Wigner, Phys. Rev. \textbf{51}, 947 (1937).

\bibitem{talmi1} I. Talmi, {\it Simple models of Complex Nuclei},
   Harwood Academic, Switzerland (1993).

\bibitem{talmip} I. Talmi and R. Thielberger, Phys. Rev \textbf{103},
923 (1956)

\bibitem{mccullen1} J.D. McCullen, B.F. Bayman and L. Zamick,
   Phys.Rev. {\bf 134}, B515 (1964);
   Technical Report NY0-9891, Princeton University.

\bibitem{iba1} A. Arima and F. Iachello, Phys Rev Lett {\bf35} (1975)
1069

\bibitem{iba2}A. Arima, T. Otsuka, F. Iachello, and I. Talmi,
Phys. Lett. \textbf{68} (1977) 205

\bibitem{iba3}J.A. Evans, J.P. Elliott, and S. Szpikowski,
Nucl. Phys. A \textbf{435} (1985) 317
\bibitem{zml1} L. Zamick, A. Mekjian, S.J. Lee, LANL preprint
nucl-th/0402089

\bibitem{zml2} L. Zamick et. al. Proceedings, 8th International Spring
Seminar on Nuclear Physics, Key Topics in Nuclear Structure, Paestum,
Italy, May 22-27, 2004

\bibitem{zamick1} L. Zamick, Phys. Rev. C {\bf 31}, 1955 (1985).

\bibitem{zamick2} L. Zamick, Phys. Rev. C {\bf 33}, 691 (1986).



\end{thebibliography}
\end{document}